\documentclass[aip,apl,preprint]{revtex4-1}
\usepackage{amsfonts}
\usepackage{amsmath}
\usepackage{amssymb}
\usepackage{graphicx}

\begin{document}

\title{Strain induced $\mathbb{Z}_2$ topological insulating state of $\beta$-As$_2$Te$_3$}

\author{Koushik Pal }
\affiliation{Chemistry and Physics of Materials Unit, Jawaharlal Nehru Centre for Advanced Scientific Research, Bangalore-560064, India}
\affiliation{Theoretical Sciences Unit, Jawaharlal Nehru Centre for Advanced Scientific Research, Bangalore-560064, India}
\author{{Umesh V. Waghmare\footnote[1]{Electronic mail:waghmare@jncasr.ac.in}}}
\affiliation{Theoretical Sciences Unit, Jawaharlal Nehru Centre for Advanced Scientific Research, Bangalore-560064, India}

\begin{abstract}
Topological insulators are  non-trivial quantum states of matter which exhibit a  gap in the electronic structure of their bulk form, but a gapless metallic electronic  
spectrum at the surface. Here, we predict a uniaxial strain induced  electronic topological transition (ETT) from a band to topological insulating state 
in the rhombohedral phase (space group: R$\bar{3}$m) of As$_2$Te$_3$ ($\beta$-As$_2$Te$_3$) through \textit{first-principles} calculations including spin-orbit coupling
within density functional theory. The ETT in $\beta$-As$_2$Te$_3$ is shown to occur at the uniaxial strain $\epsilon_{zz}$ = -0.05 ($\sigma_{zz}$=1.77 GPa), passing through 
a Weyl metallic state with a single Dirac cone in its electronic structure at the $\Gamma$ point.  We demonstrate the ETT through band inversion and reversal of parity of the top of the valence and bottom of the
conduction bands leading to change in the $\mathbb{Z}_2$ topological invariant $\nu_0$ from 0 to 1 across the transition. 
Based on its electronic structure and phonon dispersion, we propose ultra-thin films of As$_2$Te$_3$ to be
promising for use in ultra-thin stress sensors, charge pumps and thermoelectrics. 
%Our prediction is relevant to applications of epitaxial films of As$_2$Te$_3$ 
%in development of ultra-thin stress sensors, charge pumps and thermoelectrics. 
 
\end{abstract}

%\pacs{71.30.+h,71.70.Ej,71.15.Mb}

\maketitle

%\section{Introduction}
Discovery of the non-trivial electronic topology in the layered semiconductors (Bi$_2$Se$_3$, Bi$_2$Te$_3$, Sb$_2$Te$_3$) \cite{ZHANG, hseih2}
with tetradymite crystal structure  (space group: R$\bar{3}$m, No: 166)  have stimulated enormous research activity in exploration of exotic
states like  superconductivity, anomalous quantum Hall, and magneto-electric effects that have been predicted theoretically \cite{fu1, ESSIN-1, ESSIN-2}.
These materials, commonly known as topological insulators (TIs),  are  insulators in their bulk form, but exhibit a  metallic electronic spectrum at their surfaces.
The non-trivial topology of the bulk electronic states of Bi$_2$Te$_3$ type TI's arises from strong spin-orbit interactions \cite{ZHANG}. The metallic state of the surface of a 
topological insulator is protected by the time reversal symmetry, and is robust against any non-magnetic perturbations. Berry phases of electronic states at the surface
of a strong topological insulator prevent back scattering of  electrons from impurities resulting in a dissipation-less conduction of current on its surface \cite{hasan}.

Arsenic telluride has a monoclinic structure with space group C2/m ($\alpha$-As$_2$Te$_3$) at the ambient pressure, and has been investigated as a thermoelectric 
material in earlier works \cite{harman, black, yarembash, scheideantel1} showing that it has a lower thermoelectric figure of merit than Bi$_2$Te$_3$. 
There is room for improving the thermoelectric performance of As$_2$Te$_3$ by applying pressure or with epitaxial strain.
The high pressure study  of $\alpha$-As$_2$Te$_3$ by Scheidemantel et el \cite{scheideantel1} revealed a pressure induced structural phase transition from monoclinic 
($\alpha$-As$_2$Te$_3$) to rhombohedral structure ($\beta$-As$_2$Te$_3$) near 7 GPa, 
leading to dramatic enhancement in its thermoelectric power. The $\beta$-As$_2$Te$_3$ phase can 
also be synthesized by rapid quenching from 
high temperature or by compressing monoclinic $\alpha$-As$_2$Te$_3$ crystals \cite{toscani, kirkinskii1, kirkinskii2}.

The $\beta$-phase of As$_2$Te$_3$ is iso-structural to Bi$_2$Se$_3$ family of compounds with R$\bar{3}$m symmetry 
(space group No:166) having 5 atoms in the bulk unit cell.
Electronic structure of $\beta$-As$_2$Te$_3$ has been determined within a non-relativistic description \textit{i.e.} without including the spin-orbit coupling (SOC) \cite{scheideantel2}, 
and it is found to be similar to that of Bi$_2$Te$_3$ (also determined without SOC) \cite{scheideantel2} with a direct band gap of 0.12 eV at the $\Gamma$ point. 
As$_2$Te$_3$ contains a relatively light element $As$ and hence relatively weaker SOC, which can however be tuned with strain or pressure  
modifying its electronic properties. For example, a number of materials belonging to different crystal symmetries (at ambiant conditions)
have been predicted theoretically from  the quantum materials repository by using a search model based on the strain-dependent electronic structure \cite{yang}. 
Motivated by this, we determine electronic structure of $\beta$-As$_2$Te$_3$ as a function of uniaxial strain along the c-axis including SOC, and show that it
undergoes a quantum phase transition on application of a modest uniaxial stress of $\sigma_{zz}$ = 1.77 GPa to an interesting topological insulating state with 
a small gap, a property which can be exploited to make devices.

%\section{Computational details}

We use a combination of two different implementations of density functional theoretical (DFT) methods (a) the WIEN2K \cite{blaha} code which is an 
all-electron full potential linearized augmented plane wave (FP-LAPW) based technique and (b) the {\sc Quantum ESPRESSO} (QE) \cite{GIANOZZI} code which treats
only valence electrons replacing the potential of ionic core  with  a smooth pseudopotential.
To obtain total energies and eigenvalues of the electrons in a solid using the FP-LAPW methods, we use a basis set achieved by dividing
the unit cell into  non-overlapping spherical regions centered at each atom and the interstitial region. Two different types of basis sets are used in these two regions.
Plane wave basis set is used in the expansion of the electronic wave functions inside the interstitial region. It is augmented by atomic like wave functions (linear combination
of the solutions of the radial Schrödinger equation and spherical harmonics) in the space inside every atomic sphere. 
These atomic-like wave functions form the basis set inside each non-overlapping atomic sphere. We use Perdew, Burke and Ernzerhof (PBE) parametrization \cite{PERDEW} of the
exchange-correlation energy functional derived with  a generalized gradient approximation (GGA) \cite{HUA}. Spin-orbit interaction has been included through a second
variational procedure \cite{macdonald, novak}. Truncation of the plane wave expansion of  electronic wave functions inside the interstitial region is specified by a cut-off 
value of R$_{mt}$*K$_{max}$ = 7 , where R$_{mt}$ is the radius of the smallest atomic sphere (muffin-tin),  K$_{max}$ = 2.8 a.u$^{-1}$ is the plane wave cut-off vector, and charge
density is Fourier expanded up to  by G$_{max}$ = 12 Ry$^{1/2}$, where G$_{max}$ represents the maximum value of $G$ vector in the Fourier expansion. We adopt the tetrahedron 
method for sampling integrations over the Brillouin zone with a $9\times9\times9$ uniform mesh of k-vectors.

Lattice-dynamical properties are determined within the framework of self-consistent density functional perturbation theory (DFPT) as implemented within the QE code \cite{baroni}.
Since the effect of SOC is negligible  on phonon frequencies and character of the vibrational modes is unchanged  without the SOC, we determine vibrational frequencies of 
$\beta$-As$_2$Te$_3$ within a non-relativistic description. We use norm-conserving pseudopotentials and plane wave basis truncated with cut-off energies of 60 Ry and 240 Ry 
in representing of wave functions and charge density respectively. In order to calculate the phonon dispersion, force constant matrices are obtained on a 2$\times$2$\times$2 
$q$-point mesh. The dynamical matrices at arbitrary wave vectors are then obtained using Fourier interpolations.

%\section{Results}

\begin{figure}
\centering
\includegraphics[width=0.55\textwidth]{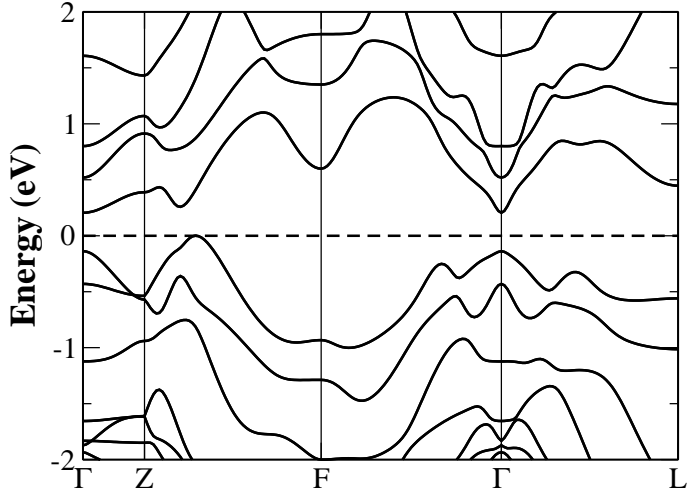}
\caption{Electronic structure of $\beta$-As$_2$Te$_3$ (space group: R$\bar{3}m$) at vanishing strain. 
Spin-orbit coupling is included in the electronic structure
calculation. The overall product of parities of the occupied bands is positive which signifies that it is a band insulator when $\epsilon_{zz}$ = 0.}
\end{figure}

Lattice parameters of $\beta$-As$_2$Te$_3$ are taken from the Materials Project repository \cite{matgenome} with a$_{hex}$=4.089 $\AA$ and c$_{hex}$= 30.306 $\AA$. 
We keep a$_{hex}$ fixed and apply uniaxial strain along the c-axis, relaxing the atomic positions at each value of the uniaxial strain until the forces on atoms become less 
that 1 mRy/bohr.  In contrast to the earlier all-electron calculation \cite{scheideantel2}, we include the SOC in determining electronic structure of $\beta$-As$_2$Te$_3$ as a function
of $\epsilon_{zz}$. From the electronic structure of  $\beta$-As$_2$Te$_3$ (see Fig. 1) at vanishing strain, 
it is clear that the valence band maxima and the conduction band minima are located at points along different directions in the Brillouin zone (i.e. band gap is indirect). However, the direct band gap at 
$\Gamma$ point is 0.35 eV, higher than the earlier estimate (0.12 eV), obtained without the SOC \cite{scheideantel2}.

Electronic states  near the Fermi level of $\beta$-As$_2$Te$_3$  are contributed largely by the $p$-orbitals of As and Te atoms.
In Bi$_2$Se$_3$-type layered materials, compressive  strain ($\epsilon_{zz}$) was found to tune the strength of the SOC by reducing the inter quintuple-layer distance \cite{young, liu}.
As $\beta$-As$_2$Te$_3$ shares similar layered crystal structure, $\epsilon_{zz}$ is expected to alter the strength of SOC and crystal field of $\beta$-As$_2$Te$_3$.  
At the compressive strain of $\epsilon_{zz}$ = -0.05, it exhibits a Weyl metallic state (see Fig. 2b), where a Dirac cone with linear dispersion (in 3-D) of the electronic bands 
appears at the $\Gamma$ point. Upon further compression of the crystal along c-axis, repulsion between the electronic 
bands due to a strong SOC leads to reopening of the bulk band gap,
accompanied by the inversion of the top of the valence and  bottom of the conduction bands at the $\Gamma$ point. Naturally, parities of the bands also change their sign 
through the band inversion. Band inversion and parity reversal of bulk electronic bands are characteristics of an electronic topological phase transition  
which has been observed in  Bi$_2$Se$_3$ (a strong $\mathbb{Z}_2$ topological insulator) as a function of strain with $\epsilon_{zz}$ = 0.06 being its critical value \cite{liu}.  
Here, we show that $\beta$-As$_2$Te$_3$ undergoes an electronic topological transition at the $\epsilon_{zz}$= -0.05, with
a uniaxial stress $\sigma_{zz}=1.77$ GPa.

\begin{figure}[!h]
\centering
\includegraphics[width=0.95\textwidth]{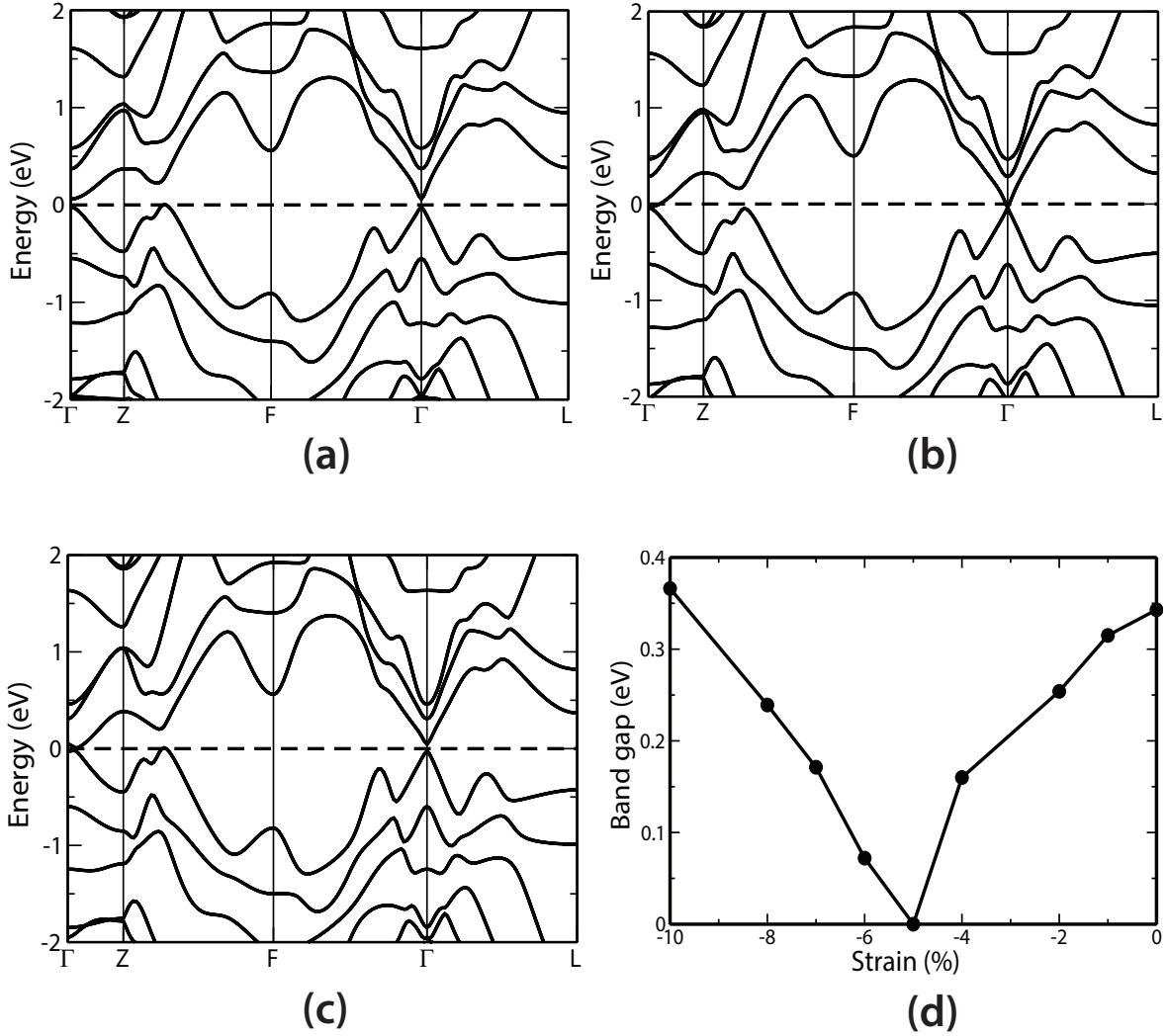}
\caption{Evolution of electronic bands  and bulk band gap of $\beta$-As$_2$Te$_3$ as a function of uniaxial strain $\epsilon_{zz}$. Electronic structures of bulk 
$\beta$-As$_2$Te$_3$ when the uniaxial strain are (a) $\epsilon_{zz}$= -0.04, (b) $\epsilon_{zz}$= -0.05, and (c) $\epsilon_{zz}$= -0.06 showing the closing and 
reopening of the bulk band gap at the $\Gamma$ point as 
a function of $\epsilon_{zz}$. (d) Variation of the direct band gap at the $\Gamma$ point as a function of uniaxial strain $\epsilon_{zz}$.
The electronic topological transition in $\beta$-As$_2$Te$_3$ passes through a metallic state at $\epsilon_{zz} =$ -0.05,  where a linearly dispersing Dirac cone 
appears at the $\Gamma$ point. The closing and reopening of the bulk band gap is accompanied with band inversion and parity reversal of states near the Fermi level. 
All the electronic structure calculations are performed taking spin-orbit coupling into account.}
\end{figure}

We now determine the $\mathbb{Z}_2$ topological invariant quantity $\nu_0$ of $\beta$-As$_2$Te$_3$ below and above the critical value of the strain using the technique of Fu and Kane \cite{fu} 
that equates the product of parities of states in the valence band manifold (see Table I)  
to $(-1)^{\nu_0}$. We find that the $\nu_0$ is 0 and 1 for $\epsilon_{zz}$ $>$ -0.05 and 
$\epsilon_{zz}$ $<$ -0.05 (the critical strain $\epsilon_{zz}$= -0.05) respectively, 
signifying that $\beta$-As$_2$Te$_3$ becomes a strong $\mathbb{Z}_2$ topological insulator for $\epsilon_{zz}$ \textless ~ -0.05. 
Similar to  Bi$_2$Se$_3$, Bi$_2$Te$_3$ and Sb$_2$Te$_3$ which are strong $\mathbb{Z}_2$ topological insulators at the ambient pressure \cite{ZHANG}, the top of valence and the bottom of conduction bands of $\beta$-As$_2$Te$_3$ 
have even and odd parities respectively in its topological insulating phase.

As shown in Fig. 2d, the band gap at the $\Gamma$ point increases with strain beyond the transition point ($\epsilon_{zz}$ \textless ~ -0.05), which is expected of a topological insulator,
but with higher value of compressive strain (\textit{e.g.} at $\epsilon_{zz} \sim$ -0.06 ), there is anti-crossing (see Fig. 2c \& Fig. 3a) of these bands along the $\Gamma$-Z direction. 
This anti-crossing behavior can be explained with group theoretical analysis of their symmetries (see the next paragraphs). While the top-most valence band touches the Fermi level
along Z-F direction, As$_2$Te$_3$ remains semiconducting at all $\epsilon_{zz} \neq$ -0.06, as evident in the electronic density of states (e-DOS) in Fig. 3b.

$\beta$-As$_2$Te$_3$ has both spatial inversion and time reversal symmetries. Inversion centre in the crystal ensures
the degeneracy of the electronic bands at \textbf{k} and \textbf{-k} i.e. $\varepsilon_{n\alpha}(\textbf{k})=\varepsilon_{n\alpha}{(\textbf{-k})}$, where 
$\varepsilon_{n\alpha}(\textbf{k})$ represents the electron energy for the n-th band with spin index  $\alpha$ at \textbf{k} wave vector in the Brillouin zone. 
On the other hand, the  time reversal symmetry implies $\varepsilon_{n\alpha}(\textbf{k})=\varepsilon_{n\bar{\alpha}}{(\textbf{-k})}$, where $\bar{\alpha}$
represents the spin opposite to $\alpha$. When both  symmetries are present, $\varepsilon_{n\alpha}(\textbf{k})=\varepsilon_{n\bar{\alpha}}{(\textbf{k})}$, \textit{i.e.} 
electronic bands acquire Kramers' double degeneracy at each \textbf{k} vector. As each electronic band in a $\mathbb{Z}_2$ topological 
insulator is doubly degenerate, the irreducible representation for each band is two dimensional (\textit{i.e. E}, according to Mulliken's symbol). 
In the Hamiltonian with SOC, the point group at any \textbf{k} vector is a double group due to inclusion of time reversal symmetry.  
The irreducible representations of  bands are hence determined by the character table of the corresponding double group of a spin-orbit coupled system \cite{dresselhaus}.  
At $\Gamma$ point (\textit{i.e.} null \textbf{k} vector) in the Brillouin zone, the group of the k-vector is $D_{3d}$, and 
electronic bands  are labeled with representations (also known as small representations) of the double group of $D_{3d}$. The top of the valence and bottom
of conduction bands  in the topological insulating state  have $E_{1/2g}$(=$\Gamma_{4+}$) and $E_{1/2u}$(=$\Gamma_{4-}$)  
symmetries respectively  (see Fig. 3a, where the scale of electronic structure has been zoomed along the $\Gamma$-Z direction) at $\Gamma$.
For \textbf{k} along z-direction ($\Gamma$-Z), the group of  \textbf{k} lacks the inversion symmetry, and therefore its  subgroup is C$_{3v}$, 
and bands along $\Gamma$-Z direction  are labeled with  irreducible representations of the double group of  C$_{3v}$.

When two bands belong to the same irreducible representation, a coupling between them is allowed by symmetry. As a result, they avoid crossing each 
other and lead to an ``anti-crossing" \cite{endo}. Electronic bands just above and below the Fermi level  along the $\Gamma$-Z direction  
anti-cross each other,  because they belong to the same irreducible representation ($E_{1/2}$=$\Gamma_4$) of C$_{3v}$.
This analysis establishes that there can be no band crossing and closure of gap along $\Gamma$-Z direction, and hence 
the electronic structure (see DOS in Fig. 3b) of $\beta$-As$_2$Te$_3$ remains semiconducting  as a function of $\epsilon_{zz}$ (including $\epsilon_{zz}$=-0.06).

As the bandgap vanishes at the electronic topological transition in $\beta$-As$_2$Te$_3$, we expect a breakdown of the adiabatic approximation in  the vicinity of the 
critical point. This broken adiabaticity would lead to Raman anomalies in a narrow range of stress near P$_c$ through a strong coupling between the electrons and phonons near the 
transition \cite{bera}. Thus, it is of fundamental importance to measure the electronic and vibrational spectra of $\beta$-As$_2$Te$_3$ as a function 
of uniaxial strain, and  confirm the presence of  electronic topological transition and associated spectroscopic anomalies in $\beta$-As$_2$Te$_3$.

Since topological insulators typically exhibit good thermoelectric properties \cite{ghaemi, fan},
we expect $\beta$-As$_2$Te$_3$ to be a better thermoelectric than its ambient pressure monoclinic phase,
consistent with the finding of Ref. [10]. 
Thin films of topological insulators like Bi$_2$Te$_3$, Bi$_2$Se$_3$ are better thermoelectric materials \cite{ghaemi} 
than their bulk counterpart due to the high mobility of the electrons on the metallic surface and low 
lattice thermal conductivity \cite{fan}. Strain engineering of thin films of Bi$_2$Se$_3$ was shown to 
be an effective way to optimize its thermoelectric figure of merit
($ZT$)\cite{saeed}, given by  $ZT$ = $\frac{\sigma S^{2}T}{\kappa}$, where $\sigma$, S and, $\kappa$ are 
electrical conductivity, Seebeck coefficient and thermal conductivity respectively. 

Low $\kappa$ is key to thermoelectric performance of a material. As acoustic phonon bands 
of $\beta$-As$_2$Te$_3$ are limited to range of frequencies less than 50 cm$^{-1}$ (see Fig. 4), 
and $\kappa$ depends quadratically on slope of the acoustic band, we expect a 
rather low thermal conductivity of $\beta$-As$_2$Te$_3$ in all the three directions. 
The narrow gap of $\beta$-As$_2$Te$_3$ will facilitate high electrical conductivity at room temperature, 
and the asymmetry in its DOS (Fig. 3b) across the gap is expected to yield a 
high S (e.g. at $\epsilon_{zz}$ = -0.06, band gap is 0.06 eV). 
Since its narrow band-gap and the symmetry of its frontier states are sensitive to uniaxial stress, 
$\beta$-As$_2$Te$_3$ has the promise of a good thermoelectric whose properties are tunable with stress field. 
%Thus, $\beta$ phase of As$_2$Te$_3$ has the potential candidate for devices based on stressed thermoelectric.

With frequencies of all its phonons less than 200 cm$^{-1}$, vibrational 
entropy gives greater stability to $\beta$-As$_2$Te$_3$ with increasing temperature. 
As the quintuple layers of $\beta$-As$_2$Te$_3$ are held together by the weak van der Waals forces, 
it can be readily prepared in the form of an ultra-thin film.   
Surface of a topological insulator exhibits a robust two dimensional electron gas (2DEG) with a high carrier mobility,
while that of a band insulator shows none. This property can be used to create a charge pump based on As$_2$Te$_3$ 
that is driven by mechanical stress field. 

\begin{figure}
\centering
\includegraphics[width=0.99\textwidth]{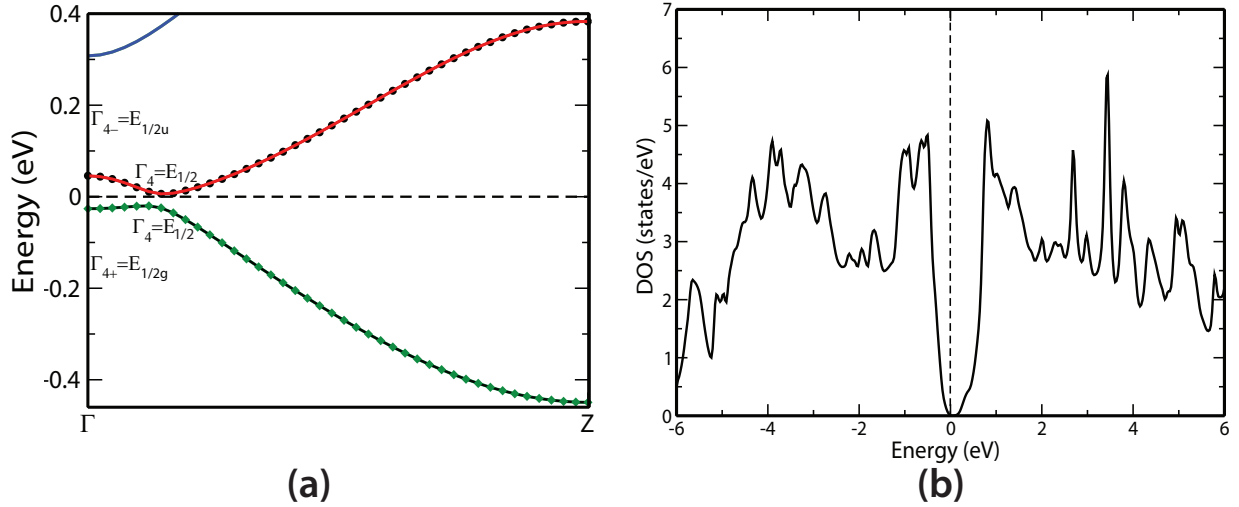}
\caption{(a) Electronic structure of $\beta$-As$_2$Te$_3$ at $\epsilon_{zz}$= -0.06 with spin-orbit coupling zoomed along the $\Gamma$-Z direction.
The electronic states near the Fermi level having the same irreducible representations lead to an anti-crossing situation as discussed in the text.   
(b) The total electronic density of states (DOS) of $\beta$-As$_2$Te$_3$ at $\epsilon_{zz}$= -0.06 show  semiconducting nature of the material in its 
topological insulating state.}
\end{figure}

\begin{table}[h]
\centering
\begin{tabular}{c|cccccccccccccccc|c}
\noalign{\smallskip} \hline \hline \noalign{\smallskip}

 $\epsilon_{zz}$ = -0.04   & + & - & + & - & + & + & - & + & - & - & + & - & - & - & ; & + & (+) \\

 $\epsilon_{zz}$ = -0.06   & + & - & + & - & + & + & - & + & - & - & + & - & - & + & ; & - & (-) \\

\noalign{\smallskip} \hline \hline \noalign{\smallskip}
\end{tabular} 

\caption {Parities of the fourteen occupied bands below the Fermi level and the lowest unoccupied band above the Fermi level across the transition
point ($\epsilon_{zz} =$ -0.05) for $\beta$-As$_2$Te$_3$.
The Product of parities of the valence band manifold are given in the rightmost column and are indicated within the brackets. Positive and negative signs within the brackets 
mean that for $\epsilon_{zz}$ \textgreater~ -0.05, $\beta$-As$_2$Te$_3$ is a band insulator which undergoes a quantum phase transition and becomes a topological insulator upon increasing the 
strain beyond it.\\}
\end{table}

\begin{figure}
\centering
\includegraphics[width=0.55\textwidth]{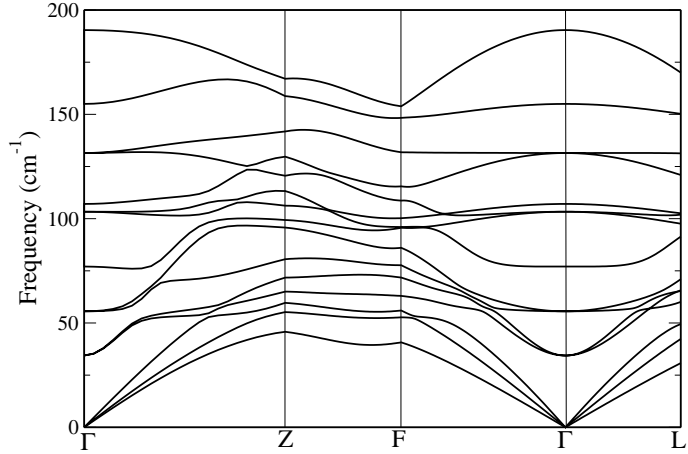}
\caption{Phonon dispersion of $\beta$-As$_2$Te$_3$ at $\epsilon_{zz}$= -0.06 calculated within a non-relativistic description.}
\end{figure}

 %\section{Conclusions}
In conclusion, we predict a uniaxial strain induced transition from band to topological insulating state 
in $\beta$-As$_2$Te$_3$  using \textit{first-principles} density functional theory based  calculations,
highlighting the importance of spin-orbit coupling. 
It exhibits a  direct band gap of 0.35 eV at the $\Gamma$ point at ambient conditions, and
passes through a Weyl metallic state with linearly dispersing bands typical of a Dirac cone at $\epsilon_{zz}$ = -0.05 
with non-zero gaps on the two sides of the transition. The ETT in the rhombohedral phase of As$_2$Te$_3$ has been 
demonstrated through the band inversion and parity reversal of the top of the valence and bottom of conduction bands 
across the critical strain, accompanied by a change in the $\mathbb{Z}_2$ topological invariant.
Finally, uniaxial stress can be used to tune electronic gap and thermoelectric performance of 
thin films of $\beta$-As$_2$Te$_3$, which augurs well for its applications.

UVW thanks funding from a JC Bose National Fellowship of the Department of Science and Technology of
Govt of India. KP thanks JNCASR for a fellowship and acknowledge the computational resources from TUE-CMS, JNCASR.

\end{document}